\documentclass{article}
\begin{document}
\title{Evidence for a hard equation of state in the cores of neutron stars}                                                                            
                                                                                                    
 \author{Chris Vuille \\ Department of Physical Sciences\\ Embry-Riddle Aeronautical 
University\\ Daytona Beach, FL 32114\\ e-mail vuille@db.erau.edu}

\maketitle

\abstract{The equation of state for matter with energy density above \newline 
$2 \times10^{14}\;\mathrm{gm/cm^3}$ 
is parametrized by $P = kN^{\Gamma}$ , where \textit{N} is the number density,
$\Gamma$  is the adiabatic index, and \textit{k} a constant. Using this scheme 
to generate thousands of models, together with data on neutron 
star masses, it is found, for a core region with constant adiabatic index, that the central density must satisfy  
$10^{15} \; \mathrm{gm/cm^3} < \rho_c < 10^{16} \;\mathrm{gm/cm^3}$, with $\Gamma>2.2$.
 Further preliminary results indicate, based on the observed 
neutrino flux from supernova 1987a, that this number must be 
considerably higher, on the order of 3.5. These results 
provide evidence for a hard equation of state in 
the cores of neutron stars.} 

\vspace{5mm}
\noindent\textbf{Author's note:} This paper was originally submitted to the Astrophysical Journal circa
1995. It is uploaded here by request of new authors who are currently exploring the use of a
 similar parameterization scheme for similar purposes. The figures and tables referred
 to in the paper are no longer available.

\section{Introduction}

There have been a number of attempts to derive the equation of 
state (EOS) of matter at high density, but all of these attempts 
have suffered under the constraint of a lack of experimental 
data. The physics of matter above about 2x10$^{14}$ grams per cubic 
centimeter is essentially unknown, and there is considerable 
controversy  \cite{Brown} as to whether or not the EOS in this 
region is hard or soft.

In view of this, a mathematical parametrization 
of the equation of state for highly compressed matter was developed which allows 
a general study of all possible equations of state. The 
process of parametrization involves several simple steps: (1) 
the choice of the mathematical relationship(s) with which to 
model the EOS; (2) the regions in which these mathematical relationships 
hold; (3) the use of continuity conditions at the boundaries 
of these regions to fix various constants.

The parameters chosen here consist of the adiabatic index (indices) 
for the matter, and the number densities at which phase transitions 
occur, if any. Alternately, energy densities may be chosen 
instead of number densities. Constraining data obtained from 
these models by physical observations may then shed some light 
on the physics in the core region.

 Using this mathematical scheme, approximately twenty-five thousand 
one-parameter models were generated for static neutron stars, 
each corresponding to a different adiabatic index in the core 
region. In addition, several derived equations of state were 
mapped onto parameter spaces of one and three dimensions, and 
the overall viability of the scheme tested by finding whether 
or not the parameter models agreed with the originals in such 
basic computations as mass and radius.

The generic parametrization is given by
\begin{equation} \label{A}
P = \kappa N^{\Gamma}
\end{equation}

where \textit{P} is the pressure, \textit{N} the number density, $\Gamma$
is the adiabatic index, and \textit{k} a constant which is determined 
by continuity. The first law of thermodynamics in the absence 
of heat can be written
\begin{equation}\label{B}
\frac{dE}{E+\frac{P}{c^2}} = \frac{dN}{N}
\end{equation}
This equation can be integrated by noticing that
\begin{equation}\label{C}
d\left(\frac{E}{N} \right) = \frac{dE}{N} - \frac{E}{N^2}dN
\end{equation}
can be suitably rearranged. Together with Equation \ref{A} and \ref{B} 
it is straight-forward to find that
\begin{equation} \label{D}
E = \frac{P}{c^2\left(\Gamma - 1 \right)}+DN
\end{equation}
where \textit{D} is a constant of integration which, like \textit{k}, is established
by continuity. (It will come as no surprise that \textit{D} has approximately 
the same magnitude as one AMU). The first number density will 
be chosen to be $N_0= 1.182 \times 10^{38}$ baryons per cubic centimeter, which 
corresponds to an energy density of $E_0 = 2.004 \times 10^{14}$ grams per cubic centimeter in the 
BPS/BBP equation of state, at the threshold of the 
nuclear regime. These numbers were obtained from tables provided 
by Jim Ipser of the University of Florida. Additional parameters 
can also be used if desired to model more complex cores involving 
phase transitions. Using more parameters also allows a better 
approximation to existing equations of state. As different regions 
are added, continuity with the previous region determines the 
constants in the parametrization for the new region.

Once the equation of state is specified, in this case Baym-Bethe-Pethick 
with the parametrization scheme implemented in the high-density 
region, it's a simple matter to integrate the Oppenheimer-Volkoff 
equations to obtain the mass and radius of the stellar model. \cite{Shapiro}

\section{Parameter Matching for Derived Equations of State}

The next step is to map existing derived equations of state onto the 
parameter space. If this is possible and yields a good approximation, 
then it can be concluded that the simple mathematical model 
can give meaningful answers to physical questions.

There are various possible choices for the optimization of the 
fit, and each choice will, in general, yield slightly different 
answers. A natural choice, for example, is to derive the equation 
of the best-fit plane in P-E-N space, subject to continuity with 
the actual equation of state at the onset of the nuclear regime. 
While this results in a fairly good mapping, a better fit can 
be obtained by ignoring the energy density E altogether and finding 
the best fit line of $\ln(P)$ to $\ln(N)$, subject to continuity, that 
is, the line must contain the point $(\ln(N_o),\ln(P_o))$, where $N_o$ 
and $P_o$ are the first fiducial density and the pressure corresponding 
to it, respectively. With the data points ranging from 0 to $L$, 
this is found to be

\begin{equation}
\Gamma = \frac{\sum_{i=1}^{L} \left(\ln(N_0) - \ln(N_i) \right
)\left(\ln(P_0 )-\ln(P_i)\right )} {\sum_{i=1}^{L} \left(\ln(N_0)-\ln(N_i)\right)^2}
\end{equation}

One, two, three or more regions can be strung together in tandem. The table
 shows the results for several representative equations of state when 
using either one adiabatic index parameter or two adiabatic parameters 
together with a transition density parameter. In the case of 
one parameter, the variational calculation immediately gives 
the best fit. In the three-parameter case, the transition number 
density was changed by small steps through the domain of definition, 
and the lowest sum of squares deviation of the natural log of 
the pressures was the criterion for selection of the best fit, 
with the pressures chosen directly off the tables provided by 
Ipser. An average percent difference was calculated by means 
of
\begin{equation}
\Delta P_{av} = \frac{1}{L} \sum_{i=1}^{L} \frac{\left|P_i - kP_i^{\Gamma}\right|}{P_i}
\end{equation}

Here, $P_{i}$ is the pressure from the table. The strength of this 
average deviation condition is its simplicity. The weakness is 
that larger pressures, where deviations may result in more serious 
perturbations of the numerical calculation, are treated on the 
same footing as smaller pressures.

An excellent test of this fitting procedure is to attempt to 
reproduce a table of masses and radii for various central densities 
obtained from a derived equation of state, such as Moskowski, 
or Baym-Bethe-Pethick. This was done for several different equations 
of state. Most of the values differed from the actual values 
by only a few percent at most. Three different Pandharapande 
models were exceptional, with deviations exceeding twenty percent 
in some cases. This was due to artificial discontinuities in 
the tables where the BBP equation of state was joined to Pandharipande's 
EOS. The discontinuity was evidenced by the appearance of a zone 
with high adiabatic index ($\Gamma ~ 5$) between the top of the BBP EOS and the first entry 
in Pandharipande's EOS. According to John Friedmann \cite{Fried2}, this 
can be fixed by adjusting an arbitrary constant in the potential 
of Pandharipande's theory. Because these tables are widely circulated, 
users of Pandharipande's equation of state should be aware 
that some adjustment in this potential may be necessary.

\section{The Equation of State Inside Neutron Stars}

Twenty-five thousand single-parameter models were generated with 
adiabatic indices ranging between 4/3 and 10/3 and central density 
between $2 \times 10^{14}$ and $1 \times 10^{16}$ grams per cubic centimeter. The 
stars were taken to be static, with only a single phase in the 
core region, described by a single adiabatic parameter. In each 
case, the mass and radius of the star was calculated. \par

The radii of neutron stars are only poorly known, though it is 
generally thought they lie between fifteen and thirty kilometers. 
Masses, on the other hand, are quite well known from measurements 
made on binary systems. Therefore, a graph of the mass of the 
star against adiabatic index of the core region and log of the 
central density was created. The results are presented in Figure 
1, which displays several mass contours projected onto the plane 
of the independent variables. Most neutron star masses fall between 
1.27 and 1.65 solar masses \cite{Thorsett}; therefore 
contour curves for these masses are plotted. In addition, the 
curve where the adiabatic speed of sound reaches the speed of 
light is shown, along with a stability curve. Models falling 
to the right of either one of these latter two lines can be rejected 
on physical grounds. \par

Altogether, these curves provide constraints to the possible 
models. Under the given assumptions, it appears that neutron 
stars for which measurements are available must have central 
densities between $1 \times10^{15}$ and $1 \times 10^{16}$ grams/cubic centimeter, 
and that the adiabatic index must be in excess of about 2.2.\par

Because neutrino flux measurements from the 1987 supernova in the 
Large Magellanic Cloud favor a soft equation of state, at least 
up to four times nuclear density, we began study of a three-parameter 
case consisting of two regions in the core, where the index in 
the lower density region is soft. Preliminary results of this 
study, still in progress, indicate that the core region must 
have a much stiffer equation of state--about 3.5--in order to 
explain the observed masses. Generally, the softer the outer 
core region, the harder the inner core.\par

A final interesting and somewhat puzzling result was obtained 
by mapping the radius of the stellar models against adiabatic 
index for constant central energy density. As might be expected, 
the radius decreases as the equation of state becomes stiffer, 
since the higher pressure effectively raises the gravitational 
mass of the star. Paradoxically, this curve comes to a minimum 
and swings back up again, an effect which becomes more pronounced 
with increasing central density. It would be interesting to find 
an explanation for this behavior.

\section{Concluding Remarks}

The most important result obtained from this study is the apparent 
hardness of the equation of state at high densities under the 
two assumptions that the core has a single phase and that rotation 
does not dramatically affect the mass of the star. This supports 
a previous and very different analysis \cite{Fried1}. Derived equations of state,however,
 are better modeled by three parameters--two indices 
and a transition number density at the point of a phase transition. 
The neutrino flux of supernova 1987a supports a soft equation 
of state up to about four times nuclear density, but preliminary 
results indicate that, in this case, there must be an even harder 
core region beyond that, with adiabatic index around 3.5. Rotation 
will probably not change the results very much, but this, too, 
should be studied. This work is currently in progress; it is 
expected that the bounds on the central density and adiabatic 
index will be narrowed somewhat. Regardless, it appears that 
the observed masses of neutron stars point to a hard equation 
of state in the core region, with a central density between 10$^{15}$ 
and 10$^{16}$ grams per cubic centimeter.

\section{Acknowledgements}

I would like to thank Jim Ipser for numerous valuable conversations, 
and for the tabulated equations of state. This work was supported 
by the NASA-JOVE program, a joint venture of NASA and Embry-Riddle 
Aeronautical University.\\

\end{document}